\documentclass{ifacconf}

\usepackage{graphicx}      
\usepackage{natbib}        
\usepackage{amsmath}
\usepackage{amssymb, mathrsfs}
\usepackage{comment}
\usepackage{url}
\begin{document}
\begin{frontmatter}

\title{Gain-Scheduling Data-Enabled Predictive Control for Nonlinear Systems with Linearized Operating Regions} 

\thanks[footnoteinfo]{Code available: \url{https://github.com/SebsDevLab/GS_DeePC.git}
}

\author[First]{Sebastian Zieglmeier} 
\author[First]{Mathias Hudoba de Badyn} 
\author[Second]{Narada D. Warakagoda}
\author[Second]{Thomas R. Krogstad}
\author[First]{Paal Engelstad}

\address[First]{Department of Technology Systems, University of Oslo, 2027 Kjeller, Norway (e-mail: {sebastiz, mathihud} @uio.no)}
\address[Second]{Norwegian Defence Research Establishment, 2027 Kjeller, Norway}

\begin{abstract}                
This paper presents a Gain-Scheduled Data-Enabled Predictive Control (GS-DeePC) framework for nonlinear systems based on multiple locally linear data representations. Instead of relying on a single global Hankel matrix, the operating range of a measurable scheduling variable is partitioned into regions, and regional Hankel matrices are constructed from persistently exciting data. To ensure smooth transitions between linearization regions and suppress region-induced chattering, composite regions are introduced, merging neighboring data sets and enabling a robust switching mechanism. The proposed method maintains the original DeePC problem structure and can achieve reduced computational complexity by requiring only short, locally informative data sequences. Extensive experiments on a nonlinear DC-motor with an unbalanced disc demonstrate the significantly improved control performance compared to standard DeePC.
\end{abstract}

\begin{keyword}
Data-driven Model Predictive Control, Gain Scheduling, Composite Regions, Nonlinear Systems. 
\end{keyword}

\end{frontmatter}

\section{Introduction}

Data-enabled Predictive Control (DeePC) was introduced in \citep{Coulson_2019} as a fully data-driven alternative to classical Model Predictive Control (MPC). Instead of relying on a parametric model that must be identified, DeePC formulates the prediction and control problem directly in terms of measured input–output trajectories. This approach eliminates the time-consuming modeling step and avoids the performance degradation that often arises from model mismatch or unmodeled dynamics. By exploiting Willems’ fundamental lemma \citep{Willems_2004}, DeePC represents all possible trajectories of a deterministic linear time-invariant (LTI) system using Hankel matrices formed from persistently exciting data, thereby offering a linear predictor without requiring explicit system identification or knowledge.
The fundamental lemma (and thus the DeePC representation) assumes a deterministic LTI system. In practical settings, however, systems are noisy and often exhibit nonlinear behavior. As shown in \citep{Coulson_2019, Elokda_2021, Huang_2023, Zieglmeier_2025, Zieglmeier_2025_Semi}, these limitations can be mitigated through appropriate regularization, which effectively relaxes the strict consistency requirements of the lemma. With suitable regularization terms, DeePC has been shown to handle significant measurement noise and nonlinearities to a certain degree, yielding robust predictive control performance. Nevertheless, when system nonlinearities become too strong, a single global linear data representation becomes insufficient, and prediction accuracy as well as closed-loop performance may degrade \citep{Naef_2025, Zieglmeier_2025}.\\
In classical control practice, nonlinear systems are frequently addressed by gain-scheduled control approaches. Here, the nonlinear system dynamics are approximated by a collection of linear models, each valid in a specific operating region \citep{Khalil_2002}. Controllers (eg., PI or linear MPC) are then designed for each region, and a scheduling mechanism interpolates or switches between them during operation. This paradigm allows nonlinear systems to be represented as a set of locally valid linear subsystems, thereby preserving both controller simplicity and closed-loop stability under suitable scheduling rules.

The same principle can be transferred to DeePC. Instead of constructing a single data representation valid for the entire system range with different nonlinear dynamics, one may construct multiple data sets, each corresponding to a local operating region, and schedule between them online. This leads to a Gain-Scheduled DeePC framework, where region-specific Hankel matrices represent the local linear dynamics, and the controller selects the appropriate data matrices according to the current operating condition. In this way, GS-DeePC inherits the advantages of gain scheduling while remaining fully data-driven and model-free.

Two related approaches address nonlinear DeePC within different frameworks. \cite{Verhoek_2021, Verhoek_2026} proposed LPV-DPC, which generalizes DeePC for linear parameter-varying (LPV) systems using measurable scheduling variables and an LPV extension of the fundamental lemma \citep{Verhoek_2025}. This enables nonlinearities to be captured within a single LPV-based data representation. The approach has a strong theoretical foundation, but in the context of nonlinear systems, successful application depends on accurate, known scheduling signals and a suitable LPV embedding, which leads to a more complex formulation of DeePC. In the numerical results of this work, LPV-DPC is used as the benchmark.\\
More recently, \cite{Naef_2025} introduced Select-DeePC, which avoids predefined regions or models by selecting data sequences that are closest to the current operating point. While this increases flexibility, it requires a more careful handling in the online data-selection process to ensure sufficiently rich data and persistence of excitation.

This work develops a systematic GS-DeePC framework with the following contributions:
\begin{itemize}
    \item A structured gain-scheduling extension of DeePC based on regional and composite Hankel matrices, including a switching logic designed to suppress chattering between neighboring regions.
    \item A practical methodology for data collection, regional partitioning, and construction of mosaic-Hankel matrices, enabling locally consistent linear data representations for nonlinear systems.
    \item Extensive nonlinear experiments and sensitivity analyses demonstrating the performance impact of the number of linearization regions.
\end{itemize}
Together, these elements establish a coherent framework for applying DeePC to nonlinear systems through scheduled, linearized data representations, merging the benefits of data-driven predictive control and classical gain-scheduled design. 

Furthermore, another approach, developed independently and published at the same time as our work, also considers region-based extensions of DeePC \citep{Guerrero_2025}. While both approaches share the idea of using multiple data sets corresponding to different operating regimes, the present work contributes in several distinct directions. First, we provide a systematic and explicit data-collection methodology, including region partitioning, persistency-of-excitation requirements, and a discussion of the trade-off between region width and data availability. Second, instead of overlap bands and parameter interpolation, we introduce composite Hankel matrices that merge neighboring regions directly at the data level, and dwell-time-based switching rules, both aimed at suppressing chattering and ensuring smooth transitions between linearization regions. Third, we emphasize the computational benefits of regional Hankel matrices and analyze how GS-DeePC reduces data requirements compared to standard DeePC. Finally, we conduct an extensive nonlinear benchmark including a sensitivity analysis over the number of linearization regions and a direct comparison with LPV-DPC. 

This paper is organized as follows. Section \ref{sec:notation} outlines the notation, and Section \ref{sec:deepc} introduces the theoretical foundations of DeePC. The proposed GS-DeePC framework, including its control structure and data-collection strategy, is presented in Section \ref{sec:GS-DeePC}.  Numerical results are discussed in Section \ref{sec:results}, including stress tests that examine the impact of the number of linearization regions. Finally, Section \ref{sec:conc} provides conclusions and directions for future work.

\section{Notation}
\label{sec:notation}
$x_k \in \mathbb{R}^{n_x}$ denotes the value of $x$ at the discrete time step $k$, where $\mathbb{R}$ denotes the set of real numbers and $n_x$ the dimension. 
$\| x\|_W^2$ denotes the squared weighted euclidean norm of the vector $x$ with the positive weighting matrix $W$, where $\| x \|_W^2 := x^\top Wx$.
A system is partitioned into $n$ number of regions with $\{M_i\}_{i=1}^{n}=[M_1,\dots,M_n]$ where $M_i$ denotes each individual region. 

\section{Data-Enabled Predictive Control} 
\label{sec:deepc}
The DeePC methodology, proposed in~\citep{Coulson_2019}, relies on sufficiently informative input–output data to capture the system dynamics implicitly. This requirement is known as persistency of excitation~\citep{Willems_2004}.  
Willems’ Fundamental Lemma states that, for any controllable and observable LTI system, every input–output trajectory can be represented as a linear combination of previously collected and persistently excited data.  
To exploit this property, the measured input and output sequences, $u_{\mathrm{data}}$ and $y_{\mathrm{data}}$, are arranged into block Hankel matrices $\mathscr{H}(u)$ and $\mathscr{H}(y)$ with $r$ block rows and $c$ columns as formalized in~\citep{Coulson_2019}. 
\begin{equation}
\mathscr{H}(u)\!=\!\left[\!
\begin{array}{@{}c@{}c@{}c@{}c@{}}
u_1 & u_2 & \cdots & u_c \\
\vdots & \vdots & \cdots & \vdots \\
u_r & \,u_{r+1} & \,\cdots & u_{r+c-1}
\end{array}
\!\right]
\label{eq:Hankelmatrix}
\end{equation}
For noise-free data, persistency of excitation corresponds to these matrices having full row rank~\citep{Coulson_2019}. When the data are noisy, a quantitative notion of excitation strength is introduced in~\citep{Coulson_2022_1}, linking the informativeness of the experiment to the singular values of the data matrix.  
In the DeePC formulation, the Hankel matrices are partitioned into past and future components as shown in~\eqref{eq:UdYd}. The matrices $(U_p, Y_p)$ encode the past system behavior used for initialization, while $(U_f, Y_f)$ represent future trajectories employed for prediction.
\begin{equation}
\mathscr{H}(u) = 
\begin{bmatrix} U_p \\ U_f \end{bmatrix},
\qquad
\mathscr{H}(y) = 
\begin{bmatrix} Y_p \\ Y_f \end{bmatrix}
\label{eq:UdYd}
\end{equation}

Consider now a discrete-time system with input $u \in \mathbb{R}^{m}$ and output $y \in \mathbb{R}^{p}$ over a prediction horizon $N$.  
The vectors $u_{\mathrm{ini}}$ and $y_{\mathrm{ini}}$ collect the most recent $T_{\mathrm{ini}}$ input–output measurements, while $u$ and $y$ denote the future inputs and predicted outputs.  
Using these quantities, DeePC formulates the finite-horizon optimization problem
\begin{equation}
\scalebox{1}{$
\begin{array}{ll}
\underset{g, u, y, \sigma}{\operatorname{min}} 
&  \|r - y \|_Q^2 + \| u \|_R^2 
+  \| g \|_{\lambda_g}^2 +  \| \sigma_y \|_{\lambda_y}^2 \\
\text {s.t.} &
\left(\begin{array}{c}
U_{{p}} \\ Y_{{p}} \\ U_{{f}} \\ Y_{{f}}
\end{array}\right) g
=
\left(\begin{array}{c}
u_{{\mathrm{ini}}} \\ y_{{\mathrm{ini}}} \\ u \\ y
\end{array}\right)
+
\left(\begin{array}{c}
0 \\ \sigma_y \\ 0 \\ 0
\end{array}\right), \\[8pt]
& u_{\text{min}} \le u \le u_{\text{max}}, \\
& y_{\text{min}} \le y \le y_{\text{max}}, \\
& u_k \in \mathcal{U}, \; y_k \in \mathcal{Y}, \quad \forall k \in \{0, \ldots, N-1\}.
\end{array}
$}
\label{eq:DeePC}
\end{equation}

Here, $g \in \mathbb{R}^{T - T_{\mathrm{ini}} - N + 1}$ serves as the coefficient vector that reconstructs admissible trajectories from the data, and $r \in \mathbb{R}^{p}$ denotes the reference signal.  
The weighting matrices $Q \in \mathbb{R}^{p \times p}$ and $R \in \mathbb{R}^{m \times m}$ are positive definite.  
Regularization terms $\| g \|_{\lambda_g}^2$ and $\| \sigma_y \|_{\lambda_y}^2$, following~\citep{Huang_2023}, enhance robustness to noise and modeling inaccuracies.  
Input and output limitations are enforced through box constraints and the admissible sets $\mathcal{U}$ and $\mathcal{Y}$.  
Robustness and stability guarantees for DeePC are established in~\citep{Berberich_2020, Berberich_2024}. 

\section{Gain-Scheduled DeePC}
\label{sec:GS-DeePC}

The GS-DeePC framework extends the classical DeePC formulation to nonlinear systems by combining multiple locally linear data representations. The operating domain of a scheduling variable is partitioned into several regions, each associated with a Hankel matrix constructed from persistently exciting data in that region.

\subsection{Construction of Regional Hankel Matrices}
\label{sec:data_collection}
The data-collection process follows four steps: (i) recording input-output trajectories together with the scheduling variable, (ii) partitioning the operating range and extracting corresponding data regions, (iii) assembling regional Hankel matrices, and (iv) constructing composite regions.

\subsubsection{Step 1: Recording Sequences}
Following the standard DeePC setup, we collect an input-output trajectory 
$\{u(k),y(k)\}_{k=1}^{T}$ and, in addition, record a scheduling variable 
$\rho(k)$ that parameterizes the dominant nonlinear behavior.

\textbf{Assumption~1:}
\label{ass:measurable_rho}
The scheduling variable $\rho(k)$ is measurable and expressible as 
\begin{equation}
    \rho(k) = \phi(x(k),u(k),y(k))
\end{equation} 
for a known mapping $\phi$.

\textbf{Assumption~2:}
\label{ass:knowledge_nonlinearity}
Sufficient system knowledge is available to select a scheduling variable that captures the primary source of nonlinearity.

\textit{Remark 1:}  
Assumptions~1–2 may be relaxed in practice. Using an output component as a surrogate scheduling variable and partitioning its range finely often yields reasonable results, but such heuristics do not guarantee that the underlying nonlinear mechanism is accurately parameterized. Domain knowledge remains important to ensure meaningful partitioning.

\subsubsection{Step 2: Extracting Regional Data}
The observed range of $\rho(k)$ is divided into $n$ regions
\begin{equation}
\{M_i\}_{i=1}^{n}=[M_1,\dots,M_n]
\end{equation}
each defined by interval boundaries with the according limits $l \in \mathbb{R}$ in \eqref{eq:boundaries}, where $l_1$ and $l_{n+1}$ are the lower and upper limits of the overall system, respectively. 
\begin{equation}
b_i := \{\, \psi \in \mathbb{R} \mid l_i \le \psi \le l_{i+1} \,\} \; \text{for} \;\; i=1,\dots,n
\label{eq:boundaries}
\end{equation}
Throughout this work, we employ a uniform partition such that all regions have equal width $w \in \mathbb{R}$ as in~\eqref{eq:width}.
\begin{equation}
    w=\frac{l_{n+1}-l_1}{n}
    \label{eq:width}
\end{equation}
For the upper boundary limits of the regions $M_i$ follows
\begin{equation}
    l_{i+1}=l_i+w \;\; \text{for} \;\; i=1,\dots,n.
\end{equation}
However, with prior knowledge of the nonlinearity, non-uniform partitions may be advantageous, e.g. using more and narrower regions in strongly nonlinear regimes and fewer in nearly linear regimes.\\
For each region $M_i$, we extract all input-output subsequences for which the scheduling variable remains within its boundary $b_i$. Only subsequences of length at least $T_{\mathrm{ini}} + N$ are extracted.

\textbf{Assumption~3:}
\label{ass:enough_data_per_region}
Each region $M_i$ contains sufficiently many data points to provide valid subsequences of length $T_{\mathrm{ini}} + N$ or longer. The region width $w$ is chosen so that the system remains within each region long enough to permit collecting such subsequences. Furthermore, multiple subsequences per region need to be collected to fill the given number of columns $c$ (see Step 3). 

Subsequences shorter than the required length are discarded. If regions are selected too narrow relative to system dynamics, the scheduling variable may traverse them too quickly to satisfy this requirement.
The usefulness of regional linearizations relies on the scheduling variable evolving slowly enough that the system remains within a region for a nontrivial duration. If $\rho(k)$ changes too rapidly relative to the sampling rate or system dynamics, the extracted subsequences become too short to capture meaningful local behavior, and the resulting Hankel matrices may not represent valid local linearizations. This requirement is implicitly enforced through Assumption~3 via the minimum subsequence length and the region width.

\subsubsection{Step 3: Construction of Regional Hankel Matrices}
For each region $M_i$, Hankel matrices $\mathscr{H}_{i}(u)$ and $\mathscr{H}_{i}(y)$ are constructed from all retained subsequences. When multiple subsequences are available, we form a mosaic-Hankel matrix in the sense of~\citep{Henk_2020} by concatenating the individual Hankel blocks, resulting in a single comprehensive Hankel matrix per region.
Although different regions may yield different sequence lengths, we impose the following for uniformity.

\textbf{Assumption~4:}
\label{ass:equal_Hankel_length}
All regional Hankel matrices are constructed with the same number of columns $c$.

This ensures matching matrix dimensions across all regions and simplifies switching between Hankel matrices during online control.

\subsubsection{Step 4: Construction of Composite regions}
To reduce sensitivity to boundary transitions, we construct composite regions $C$ by merging neighboring regions as in \eqref{eq:composites}. The data and Hankel matrices of the two constituent regions are concatenated as in \eqref{eq:Comp_Hankel}, resulting in $n_C = n -1$ composite representations that spans the composite boundaries $b_{C,i}$ over both linearization regimes with \eqref{eq:Comp_boundaries}.
\begin{equation}
C_i = M_i \cup M_{i+1} \;\; \text{for} \;\; i=1,\dots,n_C
\label{eq:composites}
\end{equation}
\begin{equation}
\begin{aligned}
    &\mathscr{H}_{C,i}(u) = \big[\mathscr{H}_i(u), \mathscr{H}_{i+1}(u)\big] &\text{for} \; i=1,\dots,n_C\\
    &\mathscr{H}_{C,i}(y) = \big[\mathscr{H}_i(y), \mathscr{H}_{i+1}(y)\big] &\text{for} \; i=1,\dots,n_C
\end{aligned}
\label{eq:Comp_Hankel}
\end{equation}
\begin{equation}
    b_{C,i}=[l_{C,i}, l_{C,i+1}] = [l_i, l_{i+2}] \; \; \; \text{for} \; i=1,\dots,n_C
    \label{eq:Comp_boundaries}
\end{equation}
The Hankel matrices of the two neighboring composite regions $C_i$ yield the linearized behavior of two regions $M_i$ and $M_{i+1}$ in equal parts, sharing one linearized local behavior with each neighboring composite region $C_{i-1}$ and $C_{i+1}$, resulting in overlapping composite regions.

Switching between Hankel matrices may introduce discontinuities in the predicted trajectories, as each regional data set represents a distinct local linear approximation of the nonlinear system. The use of composite regions mitigates these effects by sharing data from two adjacent operating regimes, ensuring that the DeePC predictor varies smoothly when the scheduling variable crosses a boundary. This reduces abrupt changes in the optimizer and improves the consistency of the closed-loop behavior. Since switching only replaces the data matrices in \eqref{eq:DeePC} and does not alter the structure of the optimization problem itself, the online formulation remains unchanged.

Feasibility of the GS-DeePC optimization problem is supported by two mechanisms. First, composite regions provide overlapping local linearizations, increasing the likelihood that the active predictor remains compatible with the initialization during transitions. Second, standard DeePC regularization through $\lambda_g$ and $\lambda_{ini}$ ensures feasibility under mild inconsistencies between data sets, weak nonlinearities, or measurement noise \citep{Coulson_2019}. Because each (composite) region is constructed from locally consistent data and satisfies Assumption~5, the GS-DeePC controller inherits the practical stability and robustness properties of DeePC within each region. \\
\textit{Remark 2:} While a formal global feasibility and stability proof under switching is beyond the scope of this work, the combination of composite regions and regularization provides a pragmatic mechanism that ensures smooth, stable closed-loop behavior in all numerical experiments.

\textbf{Assumption 5:} The nonlinear system is partitioned into a sufficiently large number of regions $n$ such that, within each region $M_i$ and each composite region $C_i$, the input-output data used to construct the corresponding Hankel matrices $\mathscr{H}_{i}(u)$, $\mathscr{H}_{i}(y)$ and $\mathscr{H}_{C,i}(u)$, $\mathscr{H}_{C,i}(y)$ is generated by a locally linear or weakly nonlinear dynamics that is well approximated by an LTI representation on that region. Consequently, the DeePC controller operating on these local Hankel matrices satisfies the theoretical assumptions required in \citep{Coulson_2019, Berberich_2020} (persistence of excitation, data consistency, and feasibility) when constrained to the region boundaries $b_i$ or $b_{C,i}$, respectively.

Assumption~5 implicitly requires that each region is chosen sufficiently small such that the underlying nonlinear dynamics can be approximated by a single locally linear behavior. If the regions are too wide, the system may exhibit different nonlinear dynamics within the same region, violating the assumption of local linearity and degrading prediction accuracy. Thus, the choice of $n$ directly determines the validity of the local linear approximation and is addressed empirically in the numerical section.
On the other hand, Assumption~3 requires that each region is sufficiently wide for the system to remain within its boundaries long enough to produce subsequences of length $T_{\mathrm{ini}}+N$. Consequently, Assumptions~3 and~5 impose opposing requirements: narrow regions improve the validity of the local linearization, whereas wider regions facilitate the collection of persistently exciting data. In practice, a balance must be struck between these two effects when selecting the region width $w$ or, equivalently, the number of regions $n$.

\textit{Remark 2:}  
\cite{Guerrero_2025} introduced regions with narrow, symmetric overlap bands at both ends. While the implementation differs, both approaches follow the same underlying idea of smoothing transitions between locally linear models by overlapping regions.

\subsection{Online Region Selection}
During online operation, the controller must determine which Hankel matrix is active based on the current scheduling variable. We consider three strategies.

\subsubsection{Region Switching}
The simplest strategy assigns each value of $\rho(k)$ to exactly one region $M_i$. At each time step, the controller selects the Hankel matrix associated with the region containing the current $\rho(k)$.  
While straightforward, this mechanism may cause undesirable rapid toggling between neighboring regions when $\rho(k)$ fluctuates near a boundary, resulting in chattering and inconsistent predictions, widely studied in classic gain scheduling control \citep{Rugh_2000}.

\subsubsection{Region Switching with Dwell Time}
To mitigate boundary chattering, a minimum dwell time $T_{\mathrm{dwell}}$ can be enforced. Once a region is selected, switching is blocked for $T_{\mathrm{dwell}}$ time steps \citep{Hou_2010}. This classical stabilization mechanism reduces switching frequency at the cost of potentially delayed transitions. To be able to follow steps throughout multiple regions, we only prohibit switching back into the former selected region. \\
\textit{Remark 3: } Introducing a dwell time necessitates choosing a small dwell time or extra stability analysis \citep{Hou_2010}. In this work, the dwell time is assumed small enough not to jeopardize stability.

\subsubsection{Composite Region Switching}

Another approach to reduce boundary–induced chattering is employing composite regions, a concept analogous to hysteresis bands used in classical gain–scheduling schemes \citep{Bett_2005}. 
Each composite region $C_i$ spans two neighboring linearization regions and is treated as containing a single valid linearization for the purpose of data-driven prediction. 
While the system state remains inside a given composite region with its composite boundary $b_{C,i}$, the corresponding composite Hankel matrices are kept active in the DeePC formulation of~\eqref{eq:DeePC}. 
A switch to the Hankel matrices of the next composite region occurs only once the state exits the current composite boundaries $b_{C,i}$. 
Because each composite region covers the union of two neighboring linearized regions, this design introduces a hysteresis-like buffer: it prevents rapid toggling near boundaries while maintaining responsiveness to genuine changes in operating conditions.

\textit{Remark 4:}  
\cite{Guerrero_2025} refers to hard and soft switching, which cover similarities to our region switching and composite-region switching, yet with a different implementation of overlapping regions. Such parallels are expected, as both mechanisms are well established in the gain-scheduling literature~\citep{Bett_2005}.

\subsection{Computational Considerations}
Switching Hankel matrices incurs negligible additional online computational
cost, since it amounts only to replacing the data matrices in \eqref{eq:DeePC} without
modifying the pre-built optimization problem, if Hankel matrices of the same size are used. The online computational complexity of the optimization problem of GS-DeePC is identical to that of standard DeePC, since the optimization problem solved at each time step has the same structure, input and output dimension and initial and future horizon. The main computational distinction arises in the construction of the Hankel matrices. When applying standard DeePC to a (strongly) nonlinear system, input–output data must be collected over the entire operating range to capture all nonlinear regimes, resulting in large Hankel matrices and accordingly the computational complexity increases \citep{Berberich_2020}. In contrast, GS-DeePC partitions the operating domain into several smaller regions, allowing each region to be represented by significantly shorter data sequences that nonetheless capture the local nonlinear behavior. This leads to substantially smaller regional Hankel matrices. Due to DeePC's computational complexity scales with the size of these matrices \citep{Berberich_2020}, GS-DeePC can achieve lower computational requirements than standard DeePC while anticipating higher prediction accuracy in each operating regime.

\section{Results}
\label{sec:results}
To validate the performance of the proposed GS-DeePC, we introduce the DC-Motor with an unbalanced disc example, representing a strong nonlinear system, and compare it with the established LPV-DPC approach. LPV-DPC was introduced in \citep{Verhoek_2021} and further detailed in \citep{Verhoek_2026} The LPV-DPC used for producing the benchmark is the exact implementation on the same nonlinear system used within \citep{Verhoek_2021}.  
Compared with the proposed GS-DeePC framework, LPV-DPC differs structurally in two key aspects. First, it relies on an LPV embedding of the nonlinear system, requiring auxiliary scheduling trajectories and product terms that increase modeling effort and computational overhead. Second, LPV-DPC incorporates the scheduling variable directly into the Hankel matrices, enabling the linearization to vary at every step of the prediction horizon. In contrast, GS-DeePC employs fixed regional Hankel matrices over the entire horizon, offering a simpler formulation and requiring only local data. Thus, LPV-DPC trades a well-defined and measurable scheduling map for within-horizon variability, while GS-DeePC emphasizes simplicity, locality of data, and practical applicability.

The system under investigation is a simulation model of a DC motor coupled to an unbalanced disc, representing a physical system with inherent nonlinear dynamics. The control objective is reference tracking of the angular position $y(t) = \theta(t)$ and the differential equation is given by~\eqref{eq:diff}, where $u$ is the controlling input voltage to the system, and $m, g, l, J, \tau, K_{\mathrm{m}}$ are the physical parameters of the system,
outlined in Table~\ref{tab:Params} along with the sampling time $T_s$.
\begin{equation}
\ddot{\theta}(t)=-\frac{m g l}{J} \sin (\theta(t))-\frac{1}{\tau} \dot{\theta}(t)+\frac{K_{\mathrm{m}}}{\tau} u(t),
\label{eq:diff}
\end{equation}
Reformulated with the substitution in~\eqref{eq:substitute} yields~\eqref{eq:statespace} with the nonlinearity clearly being pointed out via $\sin(x_2)$. Therefore, the scheduling variable is chosen as $\rho(t)=y(t)$. (In comparison, LPV-DPC uses $\rho(t)=\mathrm{sinc}(y(t))$ as scheduling variable \citep{Verhoek_2021}.)
\begin{equation}
x_1 = \dot{\theta}(t), \;\; x_2= \theta(t)
\label{eq:substitute}
\end{equation}
\begin{equation}
\left[\begin{array}{l}
\dot{x}_1 \\
\dot{x}_2
\end{array}\right]=\left[\begin{array}{c}
-1 / \tau \\
1
\end{array}\right] x_1+\left[\begin{array}{c}
-m g l /J \\
0
\end{array}\right] \sin \left(x_2\right)+\left[\begin{array}{c}
K_{\mathrm{m}} / \tau \\
0
\end{array}\right] u
\label{eq:statespace}
\end{equation}
The input is constrained to $\mathbb{U} := [-0.25, 0.25]$ according to \citep{Verhoek_2021}, while the output constraint set was increased to $\mathbb{Y} := [-\pi, \pi]$ to increase the nonlinear range of the sinus-function compared to \citep{Verhoek_2025}. 

To obtain enough and sufficiently rich data that explores the full nonlinear operating range, the system is excited by a combination of low-amplitude randomized inputs and a simple boundary-reflection mechanism. The input signal is constructed as the sum of several independently generated random binary sequence (RBS) components, which provide broadband, persistently exciting variations around an evolving state trajectory. Whenever the measured output approaches a predefined upper or lower fraction of the overall range, an additional constant bias term is added to the input, effectively “pushing’’ the trajectory back toward the other range boundary. This bang–bang-like correction induces repeated upward and downward sweeps across the full range with an overlying persistently excited multi-RBS, ensuring that all operating regions are sufficiently visited and that the resulting input–output data are suitable for constructing the regional and composite Hankel matrices used in GS-DeePC, as introduced in Section \ref{sec:GS-DeePC}.

\begin{table}[ht]
\caption{Parameter of the DC-Motor with an unbalanced disc}
\resizebox{\columnwidth}{!}{%
\begin{tabular}{l|l|l|l|l|l|l}
$m$ {[}kg{]} & $g$ {[}m $\cdot$ s$^{-2}${]} & $l$ {[}mm{]} & $J$ {[}Nm$^2${]} & $\tau$ {[}-{]} & $K_{\mathrm{m}}$ {[}-{]} & $T_s$ {[}ms{]} \\ \hline
0.07        & 9.8            & 0.42         & 2.2 $\cdot$ 10$^{-4}$      & 0.6            & 15.3         & 75
\end{tabular}%
}
\label{tab:Params}
\end{table}
To evaluate the control performance in a manner aligned with the objectives of GS-DeePC, the tracking error is separated into steady-state and transient components. Specifically, a steady-state RMSE ($RMSE_\mathrm{SS}$) is computed only over the quasi-stationary intervals $t\in[2,3], [6,7], [10,11], [14,15], [18,19] \mathrm{s}$, whereas a transient RMSE ($RMSE_\mathrm{T}$) is evaluated outside these intervals. This separation is necessary because the non-smooth step reference induces large transient errors that dominate the overall RMSE and neglect the effect of improved prediction accuracy. Since the primary benefit of GS-DeePC lies in reducing the steady-state tracking error through more accurate local linearizations, splitting the error metrics enables a clearer and more meaningful assessment of the controller’s performance. Also, the relative error $e_\mathrm{rel,SS}$ is displayed only for the same quasi-stationary intervals as for $RMSE_\mathrm{SS}$ to provide a meaningful graphic evaluation. 

A rough hyperparameter (HP) tuning was performed prior to the experiments, revealing a broad performance plateau over a wide range of values for $\lambda_g$ and $\lambda_{ini}$. Consequently, the same set of HPs listed in Table~\ref{tab:HP} was used for all regions, as these values consistently lay in the plateau and yielded robust performance across all approaches. A precise, performance-dependent HP tuning for each individual region could improve the results minimally. However, such an automated region-wise tuning procedure lies outside the scope of this work.

\begin{figure}[ht]
    \centering
    \includegraphics[width=1\linewidth]{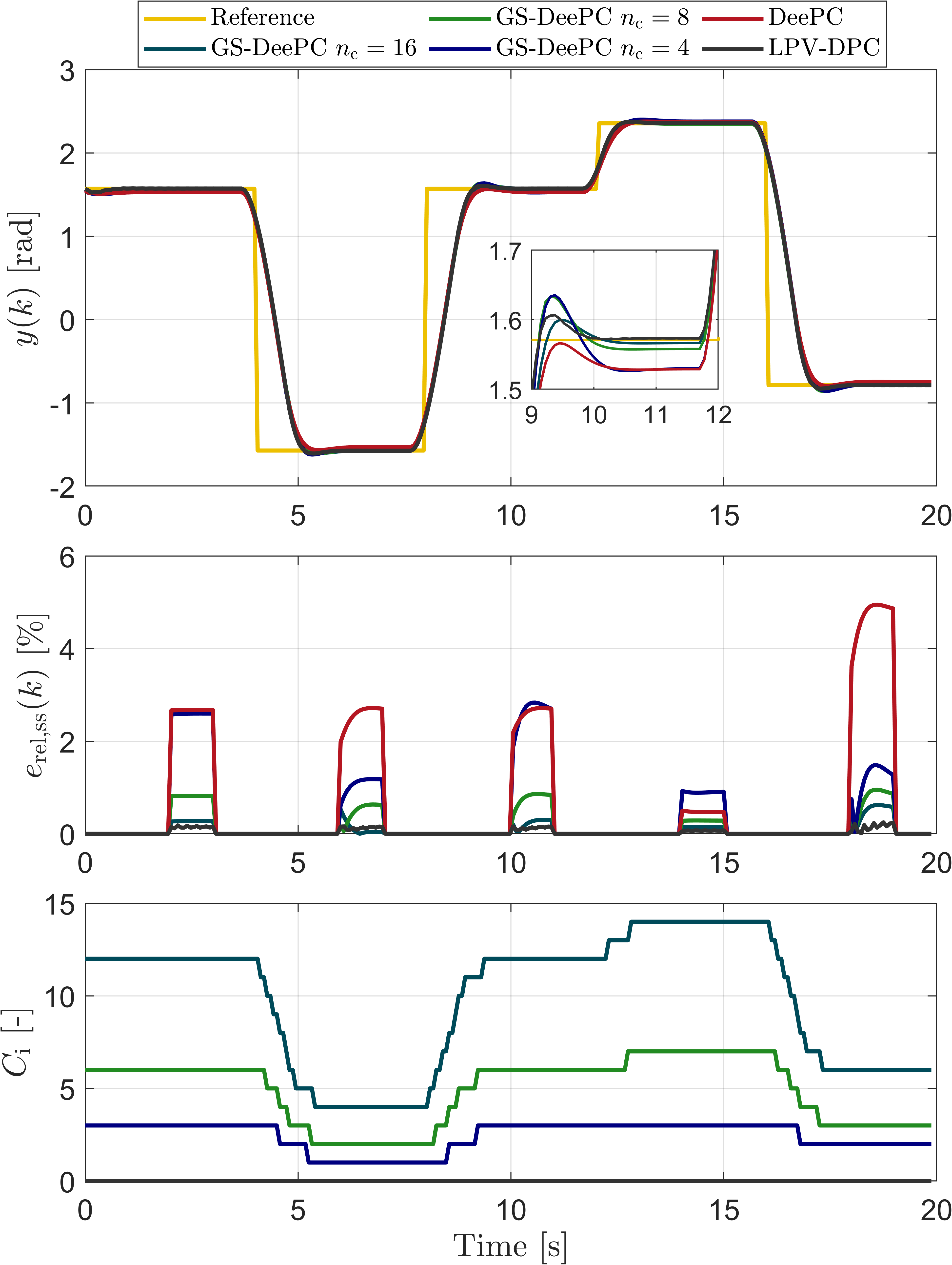}
    \caption{Reference tracking, relative state space error and active composite region for classic DeePC, LPV-DPC, and GS-DeePC with different number of linearization regions $n_{C,1}=4, n_{C,2}=8$, and $n_{C,3}=16$.}
    \label{fig:Composite_comparison}
\end{figure}

First, we analyze the results in Fig.~\ref{fig:Composite_comparison} for GS-DeePC using $n_{C,1}=4, n_{C,2}=8$, and $n_{C,3}=16$ composite regions. As the overall operating range is partitioned more finely into a larger number of locally linear regions, the prediction accuracy improves and consequently, the steady-state performance increases. This trend is clearly reflected in the relative steady-state error, where GS-DeePC consistently outperforms standard DeePC. While standard DeePC is in principle capable of controlling the nonlinear system, it does so with a substantially larger steady-state error and requires a much larger Hankel matrix to sufficiently cover the full nonlinear operating range. The HPs used for all methods are summarized in Table~\ref{tab:HP}. 
Compared to LPV-DPC, GS-DeePC with $n_{C,3}=16$ achieves nearly identical performance, with LPV-DPC yielding only marginally smaller steady-state errors. Both approaches, however, operate in a regime of comparably low steady-state error for this experiment.  
The final subplot in Fig.~\ref{fig:Composite_comparison} illustrates the online switching behavior between composite regions. The transitions are robust and stable, confirming that the composite-region construction provides a robust switching mechanism with no observable chattering.

\textit{Remark 4: }Increasing the mass $m$ scales the contribution of the nonlinear term to the overall system dynamics. 
All considered control approaches were able to control the system up to $m = 0.21\,\mathrm{kg}$. However, the performance differences between the methods became more pronounced in this regime: GS-DeePC increasingly outperformed standard DeePC, primarily due to its improved capability of handling local nonlinearities through multiple linearization regions.

\begin{figure}[b]
    \centering
    \includegraphics[width=1\linewidth]{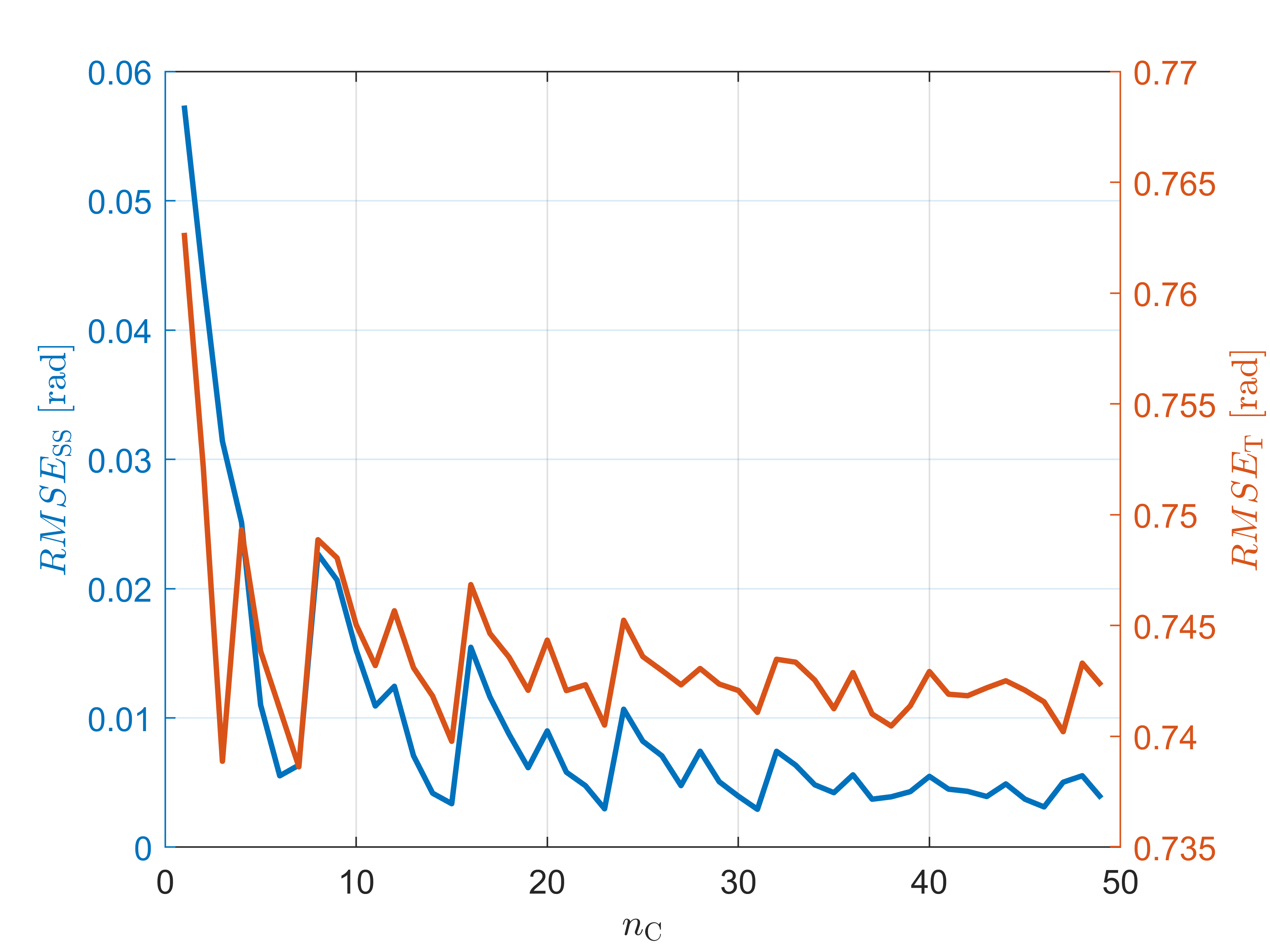}
    \caption{RMSE for the transient and steady state error intervals in dependence of composite region numbers, where $n_{C}=1$ marks classic DeePC.}
    \label{fig:RMSE}
\end{figure}

Figure~\ref{fig:RMSE} reports the steady-state and transient tracking errors as a function of the number of composite regions $n_C$. The case $n_C=1$ corresponds to standard DeePC, while the rest represent GS-DeePC with increasingly finer partitioning via the total number of composite regions $n_{C}$. Both $RMSE_\mathrm{SS}$ and $RMSE_\mathrm{T}$ improve rapidly when introducing the first few composite regions, after which the performance plateaus with small fluctuations. For $n_{C}>25$, GS-DeePC yields consistently low RMSEs, confirming that multiple local linearizations significantly enhance prediction accuracy. Increasing the number of regions also yields disadvantages as it requires using more of the collected data points of the overall data sequences, since Assumption~3 enforces a minimum subsequence length $N+T_{ini}$ for increasingly shorter regions. 

\begin{figure}[ht]
    \centering
    \includegraphics[width=1\linewidth]{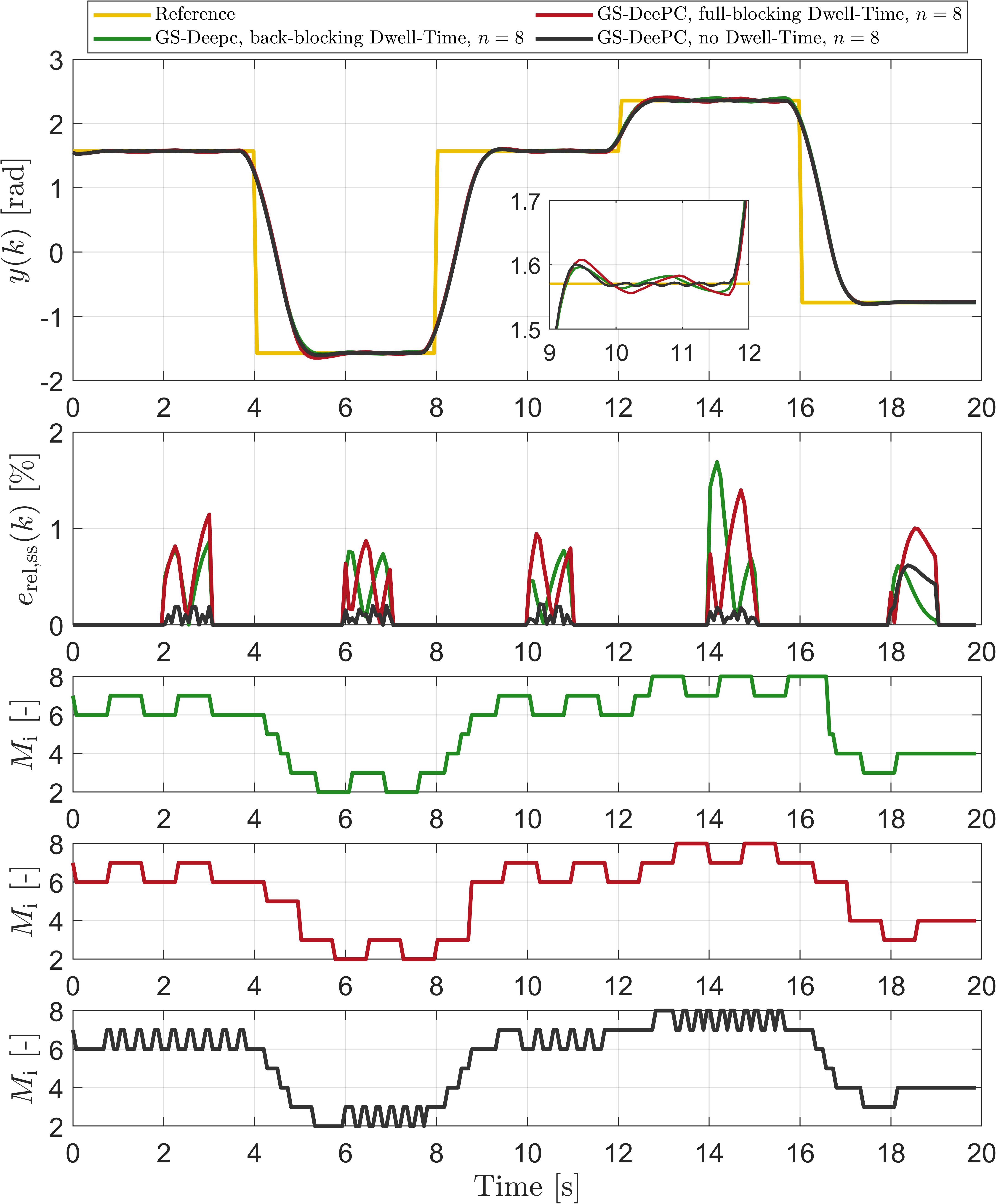}
    \caption{Reference tracking, relative state space error and active region for GS-DeePC with $n = 8$ linearization regions (non-overlapping). The dwell time $T_{\mathrm{dwell}}=10\cdot T_s$ is (i) applied to only prevent back switching, (ii) applied after every region switch, (iii) not applied.}
    \label{fig:Dwell}
\end{figure}

For the benchmark LPV-DPC implementation of \citep{Verhoek_2021}, the performance obtained in the same experiment was 
$RMSE_\mathrm{SS}=0.002$ and $RMSE_\mathrm{T}=0.74$. 
The best GS-DeePC performance was achieved at $n_{C}=23$ with $RMSE_\mathrm{SS}=0.003$ and $RMSE_\mathrm{T}=0.74$, showing that GS-DeePC approaches LPV-DPC performance despite its simpler structure. Standard DeePC, in contrast, yields significantly worse results with $RMSE_\mathrm{SS}=0.057$ and $RMSE_\mathrm{T}=0.76$, illustrating the advantage of using local linearizations over a single global data representation.
To provide further intuition for the numerical values, these RMSE metrics in Fig.~\ref{fig:RMSE} can be related to the relative steady-state errors shown in Fig.~\ref{fig:Composite_comparison}, which are based on the same experiment and illustrate the clear qualitative differences.
Figure~\ref{fig:Dwell} investigates GS-DeePC without composite regions and compares three switching strategies: (i) switching with a selective dwell rule that only prevents immediate switching back into the previously active region for $T_{\mathrm{dwell}}=10\cdot T_s$, (ii) switching with a fixed dwell time of $T_{\mathrm{dwell}}=10\cdot T_s$ blocking after every transition, and (iii) simple region switching without dwell time. All three controllers use $n=8$ (non-composite) regions and successfully stabilize the system with acceptable performance, especially when compared to standard DeePC in Fig.~\ref{fig:Composite_comparison} tracking the same reference. The reference step values were purposefully placed near region boundaries to induce chattering. Both dwell-time strategies reduce the frequency of region transitions relative to the simple switching approach without dwell time, but neither fully suppresses chattering.

Compared with the composite-region GS-DeePC in Fig.~\ref{fig:Composite_comparison}, the increased switching activity is clearly disadvantageous. The frequent region changes also lead to visible fluctuations in the system output $y$, as highlighted in the zoomed insets of both Fig.~\ref{fig:Composite_comparison} and Fig.~\ref{fig:Dwell}. These results 
confirm that composite regions provide substantially smoother transitions and more robust closed-loop behavior than dwell-time-based switching alone.

\begin{table}[h]
\caption{Hyperparameter of the different approaches employed in the experiments.}
\resizebox{\columnwidth}{!}{%
\begin{tabular}{l|c|c|c|c|c|c|c|c}
         & $N$ & $T_{ini}$ & $Q$ & $R$  & $\lambda_{ini}$       & $\lambda_g$           & $c$ & $c_{C}$ \\ \hline
GS-DeePC & 5   & 2            & 100 & 0.05 & $10^6$ & $10^3$ & 200       & 400         \\
DeePC    & 5   & 2            & 100 & 0.05 & $10^6$ & $10^3$ & 800       & -           \\
LPV-DPC  & 5   & 2            & 1   & 0.05 & -                     & -                     & 200       & -          
\end{tabular}%
}
\label{tab:HP}
\end{table}

\section{Conclusion and Further Work}
\label{sec:conc}

This paper introduced Gain-Scheduled DeePC, a practical data-driven control framework for nonlinear systems based on multiple local linear data representations. A key contribution is the use of composite regions, which share data across neighboring linearization points and thereby enable stable and smooth transitions between operating regions. 
This construction proved highly effective in practice, yielding a robust formulation that mitigates chattering and preserves prediction consistency during switching. Overall, the results show that, with a sufficiently fine partitioning of the operating domain, GS-DeePC can achieve closed-loop performance comparable to LPV-DPC while maintaining a simpler formulation. Compared to using standard DeePC for nonlinear systems, GS-DeePC benefits computationally from the reduced size of regional Hankel matrices as they need to cover shorter operating ranges.

Several directions for further work remain. First, an automated procedure for merging or splitting neighboring regions could eliminate the need for uniform partitioning and adapt the linearization grid to the underlying nonlinear dynamics more efficiently. Second, a quantitative metric for assessing the quality of each regional linearization, e.g., based on singular values or local data consistency, would support systematic region selection. Finally, fully automated region-wise HP tuning could further improve robustness and reduce performance variability across operating conditions, especially necessary for a high number of linearization regions. 
 
\section*{Declaration of Generative AI and AI Technologies in the Writing Process:}
During the preparation of this work, the author(s) used ChatGPT in order to improve the
grammar of this paper. After using this tool/service, the author(s) reviewed and edited the
content as needed and take full responsibility for the content of the publication.

\bibliography{ifacconf}            

\clearpage

\end{document}